\documentclass[11pt]{article}
\usepackage{amsfonts,amsmath,amsthm,array,amssymb}

\usepackage{graphicx}
\usepackage{picinpar,graphicx}
\usepackage{graphics}
\usepackage{multicol}
\usepackage{float}
\usepackage[sort,nocompress]{cite}
\usepackage{subcaption}

\begin{document}
\title{Population Dynamics of Wolves and Coyotes at Yellowstone National Park: Modeling Interference Competition with an Infectious Disease}

\author{Krystal Blanco$^{1}$, \\ Kamal Barley$^{2}$, Anuj Mubayi$^{2}$}
\date{}
\maketitle
\begin{center}
\footnotesize $^{1}$ Department of Mathematics, Boston University, Boston, MA\\
\footnotesize $^{2}$ Department of Applied Mathematics for the Life and Social Sciences, Arizona State University, Tempe, AZ\\
\end{center}

\begin{abstract} 
{Gray wolves were reintroduced to Yellowstone National Park (YNP) in 1995. The population initially flourished, but since 2003 the population has experience significant reductions due to factors that may include disease-induced mortality, illegal hunting, park control programs, vehicle induced deaths and intra-species aggression. Despite facing similar conditions, and interference competition with the wolves, the coyote population at YNP has persisted. In this paper we introduce an epidemiological framework that incorporates natural, human-caused and disease-induced mortality as well as interference competition between two species of predators. The outcomes generated by this theoretical framework are used to explore the impact of competition and death-induced mechanisms on predators’ coexistence. It is the hope that these results on the competitive dynamics of carnivores in Yellowstone National Park will provide park management insights that result in policies that keep the reintroduction of wolves successful.
}
\end{abstract}


\section{Motivation}
Historically, the gray wolf and coyote populations have coexisted at Yellowstone National Park (YNP) \cite{Merkle:2009aa}. However, due to their predation on farmer's livestock and the negative connotation associated to such predators, they were the focus of predator control programs in the late 1800's and early 1900's \cite{Hayward:2009aa,Fritts:1997aa}. The predator control programs utilized various methods to accelerate the extirpation of the species, including hunting, poisoning and the introduction of the parasitic mite \textit{Sarcoptic scabiei} by state veterinarians \cite{Weaver:1978aa,Merkle:2009aa}. The predator control program implemented to end both the wolf and coyote population was partially successful. Between 1914 and 1926 a total of 136 wolves were killed inside YNP \cite{Weaver:1978aa}. By 1930, wolves were completely eliminated from YNP \cite{Merkle:2009aa}. Despite similar persecution, by the end of 1930 there were 400 coyotes still present at YNP \cite{Murie:1940aa}.

In an attempt to reintroduce the now threatened and endangered gray wolf into their natural habitat and restore the original ecosystem of YNP, a total of forty-one wolves were transported from Alberta, Canada to Yellowstone National Park between 1995 and 1996 after a more than 60 year absence \cite{Smith:2007aa,Fritts:1997aa}. The reintroduction was initially successful, and the total number of wolves increased steadily with the wolf growth rate averaging about 17\% a year \cite{Hayward:2009aa} until it reached a high of 174 wolves at the end of 2003 \cite{Smith:2011aa}. However, within the last 9 years their has been a decline in the wolf population, with the northern range population experiencing a 60\% decrease in the population since 2007 and the interior range population experiencing a 23\% decrease in the same time period. The decrease in the population was rapid, and is suggestive of disease induced death \cite{Smith:2011aa}. The wolves at YNP have been affected by many diseases including canine distemper virus (CDV), canine herpes virus (CHV), canine papovirus, \textit{Brucella canis} and sarcoptic mange \cite{Smith:2007aa,Almberg:2012aa}. While all diseases have affected the wolf, in this study we focus on the effects of sarcoptic mange on the wolf population as the effects of the other diseases have been studied previously \cite{Smith:2007aa,Almberg:2010aa,Almberg:2012aa}. Moreover, since the disease was initially introduced as a control measure in 1914 for the wolf population, park management should consider treatment in the cases of extreme infection \cite{Smith:2007aa}. 

Sarcoptic mange is a highly contagious skin disease caused by the parasitic mite \textit{Sarcoptic scabiei} that burrow into the epidermis of the host species \cite{Jimenez:2010aa,Polley:2002aa}. Transmission of the the mite is caused by direct contact and contact with infected areas like dens \cite{Jimenez:2010aa}. However, the pathogens can survive off the host for days and sometimes weeks under certain microclimate conditions at the drop-off site \cite{Arlian:1989aa}. The wolves experience an allergic response to the waste secreted by the mites which causes irritation and pruritis, and leaves the infected animals suffering from alopecia, hyperkeratosis, seborrhea, scabs, ulcerations and lesions \cite{Jimenez:2010aa,Almberg:2009aa}. In severe cases it can affect the hosts entire health, leading to poor body conditions and leaving the susceptible to secondary infections or hypothermia in the winter due to the hair loss \cite{Jimenez:2010aa}. Moreover, some research suggests that wolves suffering from sarcoptic mange may change their social behavior. The weak and afflicted wolves are observed choosing to leave their pack and traveling alone; they are unlikely to survive, especially in the winter \cite{Jimenez:2010aa,Wydeven:2003aa}. Sarcoptic mange was first observed for the reintroduced wolves at YNP in 2003 when a wolf was sited at Daly Creek with hair loss \cite{Smith:2004aa}. Some populations can survive a sarcoptic mange epizootic, like the coyote. It is important to note that while coyotes do survive the epizootic, sarcoptic mange does reduce ovulation and pregnancy rates in coyotes, as well as increasing mortality rates by $\sim$70\% \cite{Pence:1994aa}. However, the effects on the population dynamics of gray wolves should be studied and monitored closely since it is considered a threat to small, recovering populations \cite{Almberg:2012aa}.

Another major threat to the recovering gray wolf population are humans. Although gray wolves are not hunted within the park, they have still been affected by vehicular death, park management actions, legal kills (i.e., if a wolf has killed a farmers livestock) and illegal hunting. Between 1995 and 2003, 38\% of reported wolf deaths were human related at Yellowstone \cite{Smith:2003ab}. Because of the high number of deaths attributed to human activity, human-caused mortality needs to be considered in order to understand the population dynamics of wolves. 

The goal of the study is to understand how different factors may affect the decline in population size of a dominant predator in relation to human-related mortality, disease, and other factors and so we consider a subordinate predator, the coyote. The sympatric predator, the coyote, at Yellowstone National Park has thrived despite facing similar persecution and environmental factors \cite{Merkle:2009aa}. We develop and analyze a mathematical model which considers the effects of disease on two competing species that are affected by a host-specific disease, and that experience human-related mortality in order to gain insight on why the dominant predator in the YNP ecosystem, the wolf, is seemingly unable to sustain a stable population while the less dominant predator, the coyote, has thrived \cite{Merkle:2009aa,Berger:2007aa}. Because wolves and coyotes compete for ungulate carcasses and habitat \cite{Merkle:2009aa}, we consider interference competition between the species in our model. 

Several models have informed the development of the framework presented in this paper. A two competing species with an infectious disease has been developed by both Han and Pugliese \cite{Han:2009aa}, and by Saenz and Hethcote \cite{Saenz:2006aa}. Our model differs from both of these models in multiple ways, including the inclusion of human related mortality. It is markedly different from the model developed by Han and Pugliese in that we do not assume competition acts only upon the death rate, but also upon the birth rate. We also refrain from considering interspecies transmission of disease as the two previous models do, since research has suggested that \textit{S. scabiei} shows high degree of host-specificity \cite{Pence:2002aa}; moreover, for social animals, like wolves and coyotes, intraspecies transmission will likely be higher than interspecies transmission \cite{Almberg:2012aa} and can be assumed negligible.  

Since the reintroduction of gray wolves has benefited the ecosystem by, for example, regulating the size of various species of ungulates and coyotes \cite{Fortin:2005aa,Switalski:2003aa,Berger:2007aa}, it is important to have continued success of the reintroduction. By studying how different factors could lead to extinction of a dominant predator in an ecosystem, we hope to provide insight to park management about some factors that could be contributing to the decline of gray wolves at Yellowstone National Park.  

\section{Competing Species with Infectious Disease and Human-Related Mortality Model}
We consider a two competing species based on the competing species with infectious disease developed by Han and Pugliese, and by Saenz and Hethcote, with the addition of human-caused mortality. It is important to note that death of coyotes by wolves occurs, mortality due to wolves is low \cite{Berger:2007aa}, and thus is not included in our model. Moreover, we assume a disease which displays host-specificity, since our motivation is the \textit{S. scabiei} mite which displays host-specificity \cite{Pence:2002aa}. Therefore we assume no interspecies transmission of the disease. We also assume that the disease has no affect on birth rates. Although research has shown a reduction in reproduction as a result of sarcoptic mange for coyotes \cite{Pence:1994aa} and for wolves \cite{Smith:2009aa}, we reduce the complexity of our model by initially assuming no effect of mange on reproduction or pup survival. Also, while some research suggests that some mammals may develop temporary immunity from sarcoptic mange \cite{Pence:1994aa,Polley:2002aa}, no conclusive argument has been reached on the existence or length of this immunity \cite{Pence:2002aa}, and we therefore exclude a recovery class from our model. Thus we model the progression of the disease using an SIS approach for each species. The coupled model is listed below:

\begin{align}
\frac{dS_1}{dt} &= \alpha_1 N_1 \left(1-\frac{N_1+\omega_{12}N_2}{k_1}\right) + \gamma_1 I_1 - \eta_1 S_1 - \beta_{1} S_1 I_1 \nonumber\\
\frac{dI_1}{dt} &=  \beta_{1} S_1 I_1 - G_1 I_1 \nonumber \\
\frac{dS_2}{dt} &= \alpha_2 N_2 \left(1-\frac{N_2+\omega_{21}N_1}{k_2}\right) + \gamma_2 I_2 - \eta_2 S_2 - \beta_{2} S_2 I_2 \nonumber \\
\frac{dI_2}{dt} &= \beta_{2} S_2 I_2 - G_2 I_2 
\end{align}

\noindent where

\begin{align}
G_1 &= \eta_1 + \delta_1 + \gamma_1 \nonumber \\
G_2 &= \eta_2 + \delta_2 + \gamma_2 
\end{align}

\begin{center}
\begin{figure}[H]
\includegraphics[scale = 0.65]{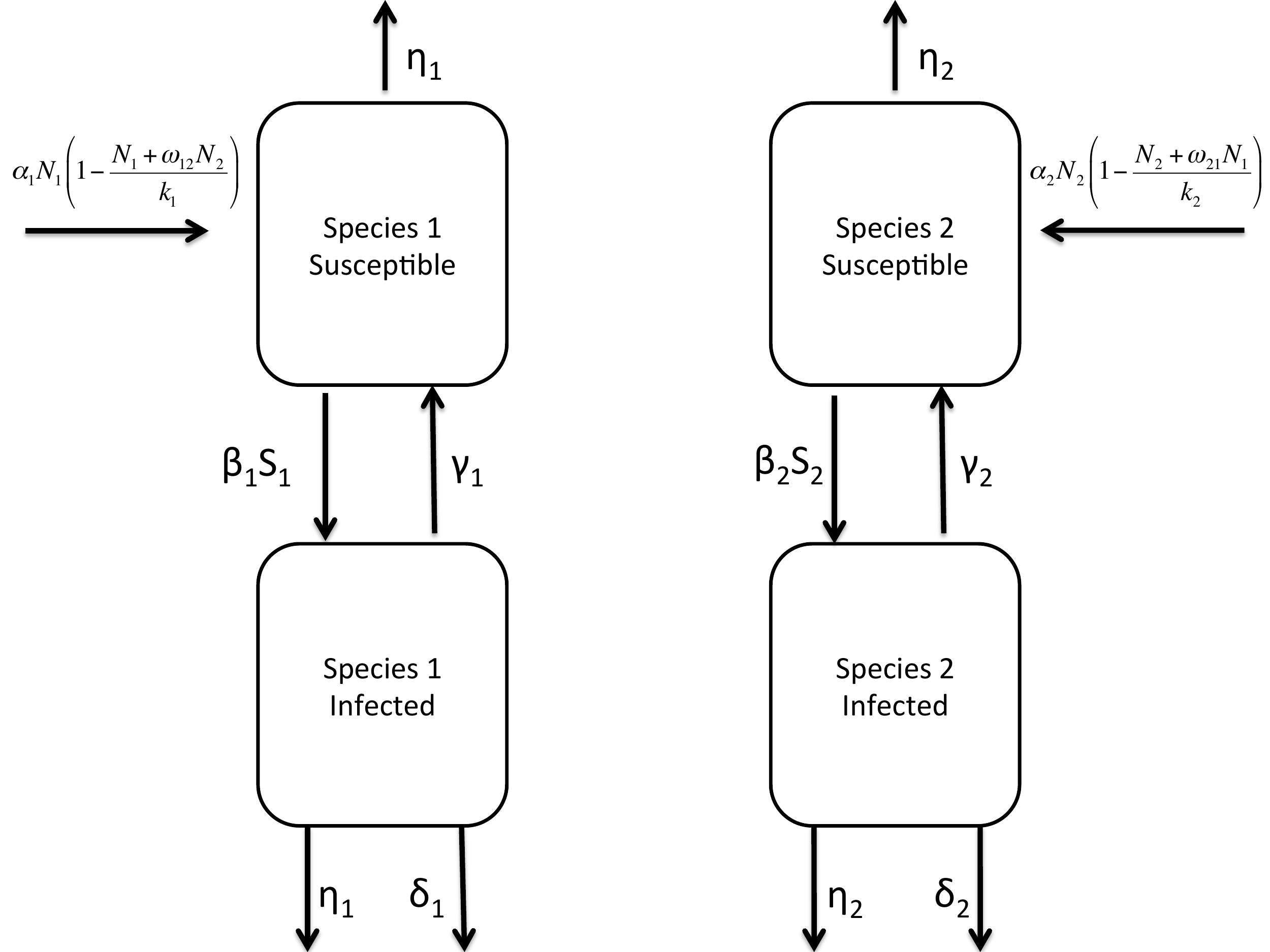} 
\caption{Compartmental Model}
\label{fig:compartmental_model}
\end{figure}
\end{center}

A compartmental diagram of the model is shown in Figure \ref{fig:compartmental_model}. The total population density of species 1 is given is given by $N_1 = S_1 + I_1$. The species 1 grows with an intrinsic growth rate $\alpha_1$, and is limited by a carrying capacity $k_1$. The inhibitory affect of species 1 on themselves is represented by $\frac{1}{k_1}$, and the inhibitory affect of species 2 on species 1 is represented by $\frac{\omega_{12}}{k_1}$. The competition coefficient $\omega_{12}$ is defined as the degree to which an individual of one species affects, through competition, the growth of the second species \cite{Schoener:1974aa}. Species 1 dies from human activities at a per capita rate $\eta_1$, and are infected by disease with a transmission potential $\beta_1$. They  recover from the disease with rate $\gamma_1$ and die from the disease at a rate $\delta_1$. Analogous classes and parameters exist for species 2. A summary of the class and parameter definitions, and their values, are given in Table \ref{parameter_definitions}.  

\begin{table}[H]
\centering
\caption{Parameters and Classes} 
\label{parameter_definitions}
\begin{tabular}{|l|l|l|}
\hline
Class & Definition &                                                                                          \\
\hline 
$N_1$           & Total population of species 1 &      \\
$N_2$           & Total population of species 2  &                    \\
$S_1$           & Susceptibles of species 1 &                                                               \\
$S_2$           & Susceptibles of species 2  &                                                             \\
$I_1$           & Infected of species 1 &                                                                  \\
$I_2$           & Infected of species 2  &                                                               \\
\hline
Parameter & Definition & Unit                                                                        \\
\hline
$k_1$           & Species 1 carrying capacity & species 1                  \\
$k_2$           & Species 2 carrying capacity & species 2 \\
$\alpha_1$      & Intrinsic growth rate of species 1 & species 1/time          \\
$\alpha_2$      & Intrinsic growth rate of species 2 & species 2/time\\
$\omega_{12}$   & Competition coefficient & species 1/species 2 \\
$\omega_{21}$   & Competition coefficient & species 2/species 1 \\
$\gamma_1$      & Per capita recovery rate for species 1 & 1/time                                         \\
$\gamma_2$      & Per capita recovery rate for species 2 & 1/time                  \\
$\eta_1$        & Per capita death rate of species 1 by humans & 1/time   \\
$\eta_2$        & Per capita death rate of species 2 by humans & 1/time  \\
$\delta_1$      & Per capita disease death rate of species 1 & 1/time  \\
$\delta_2$      & Per capita disease death rate of species 2 & 1/time     \\
$\beta_{1}$     & Transmission coefficient for species 1 & 1/time        \\
$\beta_{2}$     & Transmission coefficient for species 2 & 1/time              \\
\hline                              
\end{tabular}
\end{table}

\section{Analysis of Competing Two Predators Model in the Presence of Disease}

Using the Next-Generation Matrix \cite{Diekmann:1990aa}, we find that the basic reproduction number $R_0$ for the entire system is

$$R_0 = max (R_1, R_2)$$

\noindent where $R_1$ is the basic reproduction number for species 1 and $R_2$ is the basic reproduction number for species 2 and are defined to be

\begin{align*}
R_1 = \frac{\beta_1 k_1 (\alpha_1 - \eta_1)}{G_1 \alpha_1 } \\
R_2 = \frac{\beta_2 k_2 (\alpha_2 - \eta_2)}{G_2 \alpha_2 }
\end{align*}

Analyzing $R_1$, we see that the threshold for species 1 depends on the average infectious period for species 1, $\frac{\beta_1}{G_1}$ multiplied by the number of susceptibles we have at equilibrium when there is no infection. We arrive at similar conclusions if we analyze $R_2$. We note that because the basic reproduction number cannot be negative, we impose the restriction on our system that $G_1 < \alpha_1$ and $G_2 < \alpha_2$, i.e. that the number of species hunted is not greater than the intrinsic growth rate of the species. 

Setting the right hand side of the system to zero we found that there are at least five equilibrium points, namely extinction state $(E_0)$, two one-host disease free states $(E_1,E_2)$, two one-host endemic states $(E_3,E_4)$ and a two-host disease free state $(E_5)$. We found coexistence endemic equilibrium to be algebraically intractable.

\subsection{Trivial Equilibrium Point}
The trivial equilibrium point is 

$$E_0 = (0,0,0,0)$$ 

\noindent Two of the eigenvalues of the Jacobian $J$ of the system evaluated at $E_0$ are always negative ($-\eta_1$ and $-\eta_c$). The other two eigenvalues are negative when the following inequalities hold:

$$\alpha_1 < \eta_1 \text{ and } \alpha_2 < \eta_2$$

\noindent However, this cannot be true because the basic reproduction number would then be negative, which is not possible. Therefore the trivial solution is locally asymptotic unstable.

\subsection{One-Host Disease Free Equilibrium and Stability}
\noindent For each of the species there is a one-host disease free equilibrium point. These two equilibrium points are

\begin{align}
E_1 &= \left(\frac{k_1}{{\alpha_1} }\left(\alpha_1 - \eta_1\right),0,0,0 \right) \nonumber \\
E_2 &= \left(0,0,\frac{k_2}{\alpha_2} \left(\alpha_2 - \eta_2 \right),0\right) \nonumber 
\end{align}

\noindent For $E_1$ to exist, we need $S^{*}_1 > 0$. So we need 

\begin{align*}
\frac{k_1}{\alpha_1} \left(\alpha_1 - \eta_1 \right) > 0
\end{align*} 

\noindent which is always true. Two of the eigenvalues of the Jacobian evaluated at $E_1$ are always negative ($-G_2$ and $(\eta_1 - \alpha_1)$). Hence we need the other two eigenvalues to be negative to have local asymptotic stability; the inequalities are as follows: 

\begin{align}
-G_1 + \frac{\beta_1 k_1}{\alpha_1} \left(\alpha_1 - \eta_1 \right) &< 0 \label{eqpt1:eigenvalue1} \\
\frac{\alpha_2 k_1 \omega_{21} (\eta_1 - \alpha_1) + \alpha_1 k_2 (\alpha_2 - \eta_2)}{\alpha_1 k_2} &<0 \label{eqpt1:eigenvalue2}
\end{align}

\noindent Rearranging inequality \eqref{eqpt1:eigenvalue1}, we get that 

\begin{align*}
\frac{k_1}{\alpha_1} \left(\alpha_1 - \eta_1 \right) &< \frac{G_1}{\beta_1} \\
R_1 &< 1
\end{align*}

\noindent If $R_1 < 1$ then the disease will die out in species 1. We derive the second condition for stability from inequality \eqref{eqpt1:eigenvalue2} as follows: 

\begin{align*}
\frac{k_1}{\alpha_1} (\eta_1 - \alpha_1) \cdot \frac{\alpha_2 \omega_{21}}{k_2} + (\alpha_2 - \eta_2) &< 0 \\
\frac{k_2 (\alpha_2 - \eta_2)}{\alpha_2 \omega_{21}} &< S^{*}_1 
\end{align*}

\noindent which means that the number of susceptibles of species that remain when there is no infection must be greater than some fraction of the number of susceptibles of species 2. We can continue to arrange the above inequality so that

\begin{align*}
\frac{\beta_1 k_2 (\alpha_2 - \eta_2)}{G_1 \alpha_2 \omega_{21}} &< R_1 \\
\frac{\beta_1 k_2 (\alpha_2 - \eta_2)}{G_1 \alpha_2 \omega_{21}} \cdot \frac{\beta_2}{G_2} \cdot \frac{G_2}{\beta_2} &< R_1 \\
\frac{\beta_1 G_2}{G_1 \beta_2 \omega{21}} R_2 &< R_1
\end{align*}

\noindent Uniting these conditions we get that $E_1$ is locally asymptotically stable if 

$$\frac{\beta_1 G_2}{G_1 \beta_2 \omega_{21}} R_2 < R_1 < 1$$

\noindent Similar conditions can be defined for $E_2$.

\subsection{One-Host Endemic Equilibrium}
There are two one-host endemic equilibrium points, one for each of the species in our system. These two equilibrium points are as follows:
\begin{align*}
E_3 &= \left(\frac{G_1}{\beta_1},I^{*}_1,0,0 \right)  \\
E_4 &= \left(0,0,\frac{G_2}{\beta_2},I^{*}_2 \right)  
\end{align*}

\noindent where 

\begin{align}
A &= 1/2\,{\frac {-k_{{1}} \left( G_{1}-\alpha_{{1}}-\gamma_{{1}} \right) 
\beta_{{1}}-2\,G_{1}\,\alpha_{{1}} \pm \sqrt {-4\,\beta_{{1}} \left( -1/4
\,k_{{1}} \left( G_{1}-\alpha_{{1}}-\gamma_{{1}} \right) ^{2}\beta_{{1
}}+G_{1}\,\alpha_{{1}} \left( -\delta_1 \right)  
\right) k_{{1}}}}{\alpha_{{1}}\beta_{{1}}}} \label{A}\\
B &= 1/2\,{\frac {-k_{{2}} \left( G_{2}-\alpha_{{2}}-\gamma_{{2}} \right) 
\beta_{{2}}-2\,G_{2}\,\alpha_{{2}} \pm \sqrt {-4\,\beta_{{2}} \left( -1/4
\,k_{{2}} \left( G_{2}-\alpha_{{2}}-\gamma_{{2}} \right) ^{2}\beta_{{2
}}+G_{2}\,\alpha_{{2}} \left( -\delta_2 \right) 
\right) k_{{2}}}}{\alpha_{{2}}\beta_{{2}}}} \label{B}
\end{align}

\noindent and $I^{*}_1$ is the positive value of $A$ and $I^{*}_2$ is the positive value of $B$. We have shown in Appendix \ref{unique_pos} that only one of the values of $A$ or $B$ is positive at any given time. Because the stability of $E_3$ and $E_4$ is algebraically intractable, we have run numerical simulations to determine the behavior of our system around these equilibrium points and have found them to be locally stable.

\subsection{Two-Host Disease Free Equilibrium }
Our two-host disease free equilibrium is as follows:

$$E_5 = \left( \frac{\alpha_1 k_2 \omega_{12} (\eta_2 - \alpha_2) + \alpha_2 k_1 (\alpha_1 - \eta_1)}{\alpha_2 \alpha_1 (1-\omega_{12} \omega_{21})}, 0, \frac{\alpha_2 k_1 \omega_{21} (\eta_1 - \alpha_1) + \alpha_1 k_2 (\alpha_2 - \eta_2)}{\alpha_2 \alpha_1 (1-\omega_{12} \omega_{21})},0 \right) $$

\noindent and will exist if

$$\eta_1 <\alpha_1\text{ and }\eta_2 < \alpha_2 \text{ and } \omega_{12}\omega_{21} < 1,$$ \\

\noindent which means that the intrinsic growth rate of each population is greater than the respective rates at which they are hunted, and if the competition of the species is relatively low, both species cannot be using extreme amount of resources. We are unable to determine conditions for stability using either the eigenvalues of the Jacobian evaluated at $E_5$ or the Routh-Hurwitz criterion (see Appendix \ref{app:one-host_DFE}). We have run numerical simulations to determine the stability around the two-host disease free equilibrium and have found it locally stable.    

\subsection{Coexistent Endemic Equilibrium}
The expression of the endemic equilibrium is algebraically intractable. It is possible to have several endemic equilibrium with mass action disease transmission \cite{Bokil:2010aa}. Therefore we use numerical simulations to determine the behavior of the system if it approaches the coexistent endemic equilibrium.

\section{Numerical Simulations}
In order to understand the competitive dynamics of predators, we assume the dominance of one species over the other. In these scenarios species 1 is the dominant predator, and species 2 is the subordinate predator. 

\begin{figure}[H]
\centering
\begin{subfigure}{.5\textwidth}
  \centering
  \includegraphics[scale = 0.7]{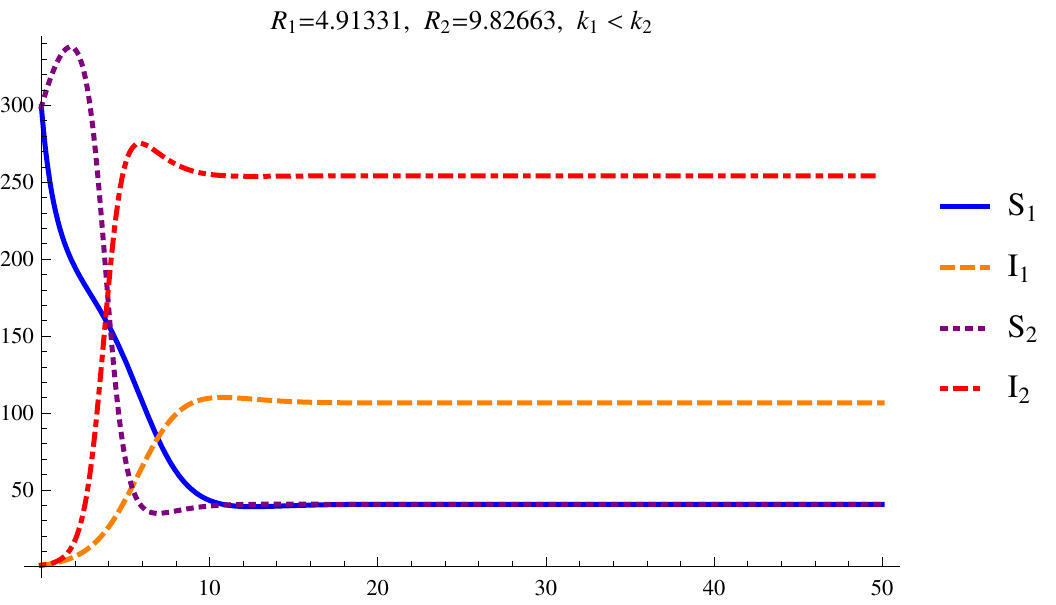}
\end{subfigure}%
\begin{subfigure}{.5\textwidth}
  \centering
  \includegraphics[scale = 0.7]{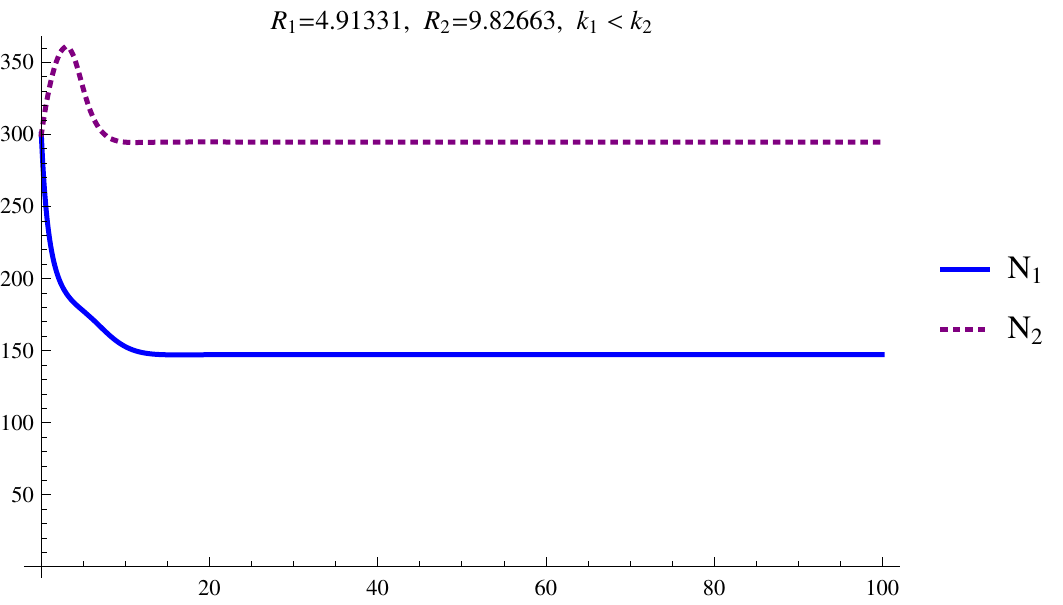}
\end{subfigure}
\caption{$\omega_{21} = 2 \cdot \omega_{12}$ and $k_2 = 2 \cdot k_1$, other parameters fixed}
\label{fig:scenario1}
\end{figure}
Although species 1 is the dominant predator, because the land can support more of the subordinate predators (its carrying capacity is larger) it's population will grow and stabilize at larger numbers and limit the growth of the dominant predator.

\begin{figure}[H]
\centering
\begin{subfigure}{.5\textwidth}
  \centering
  \includegraphics[scale = 0.7]{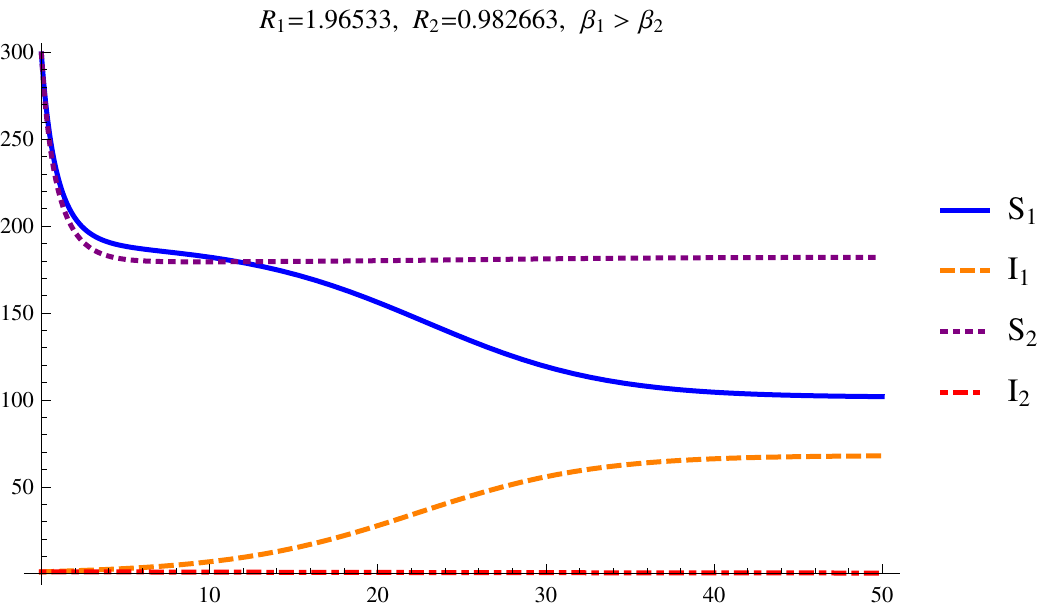}
\end{subfigure}%
\begin{subfigure}{.5\textwidth}
  \centering
  \includegraphics[scale = 0.7]{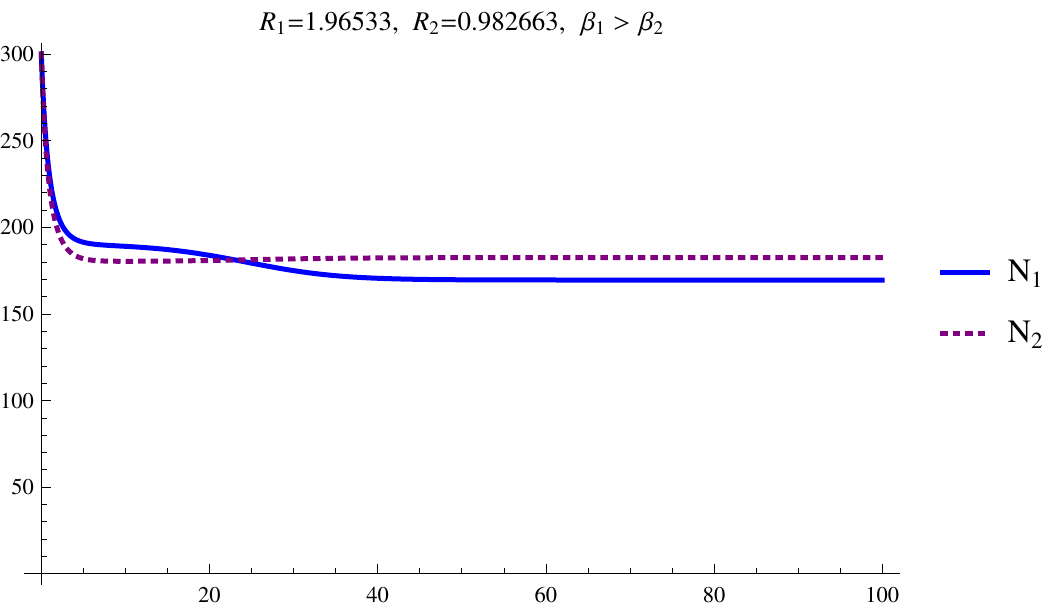}
\end{subfigure}
\caption{$\omega_{21} = 2 \cdot \omega_{12}$ and $\beta_1 = 2 \cdot \beta_2$, other parameters fixed}
\label{fig:scenario2}
\end{figure}
Although species 1 is the dominant predator, because it's infection rate is twice as high (perhaps because it is a more social creature the subordinate predator's carrying capacity is larger it's population will grow and stabilize at larger numbers than that of the dominant predator.

\begin{figure}[H]
\centering
\begin{subfigure}{.5\textwidth}
  \centering
  \includegraphics[scale = 0.7]{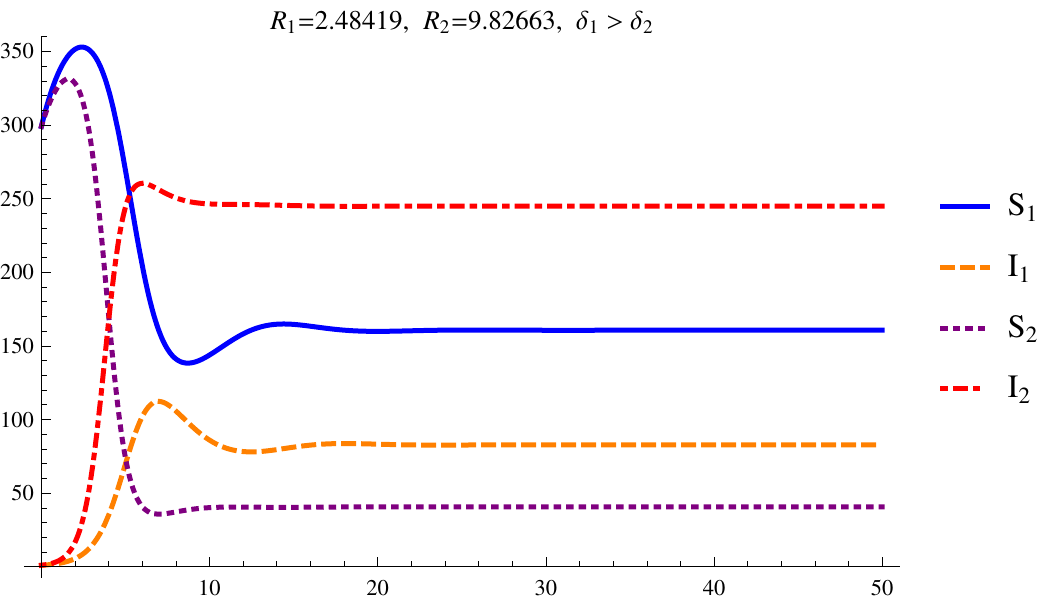}
\end{subfigure}%
\begin{subfigure}{.5\textwidth}
  \centering
  \includegraphics[scale = 0.7]{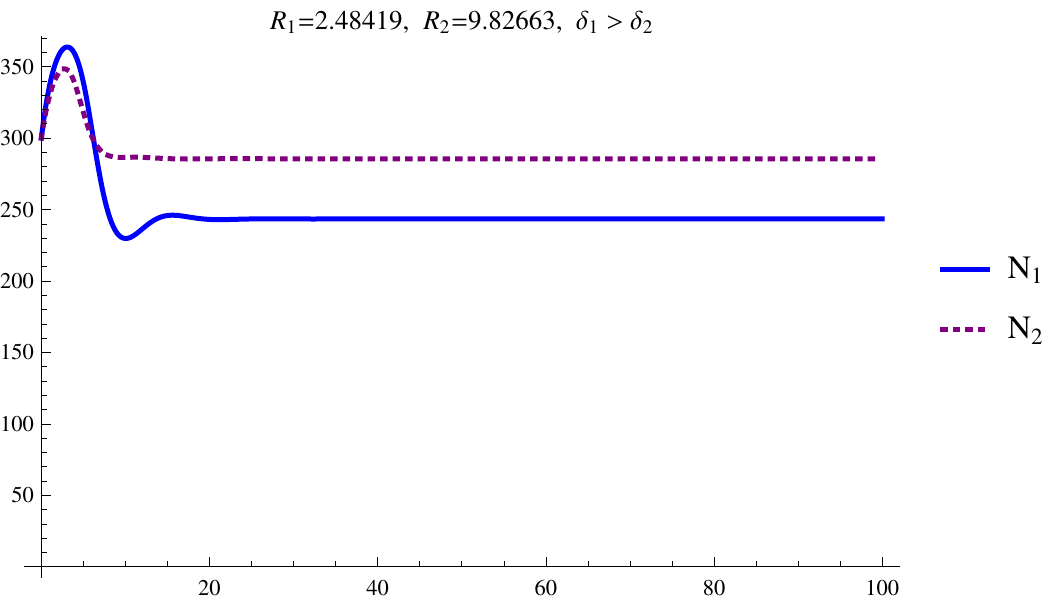}
\end{subfigure}
\caption{$\omega_{21} = 2 \cdot \omega_{12}$ and $\delta_1 = 3 \cdot \delta_2$, other parameters fixed}
\label{fig:scenario3}
\end{figure}
Although species 1 is the dominant predator, because it's disease-related death rate is three times as high (perhaps because it is more prone to secondary infections) the subordinate predator's population will grow and stabilize at larger numbers than that of the dominant predator.

\begin{figure}[H]
\centering
\begin{subfigure}{.5\textwidth}
  \centering
  \includegraphics[scale = 0.7]{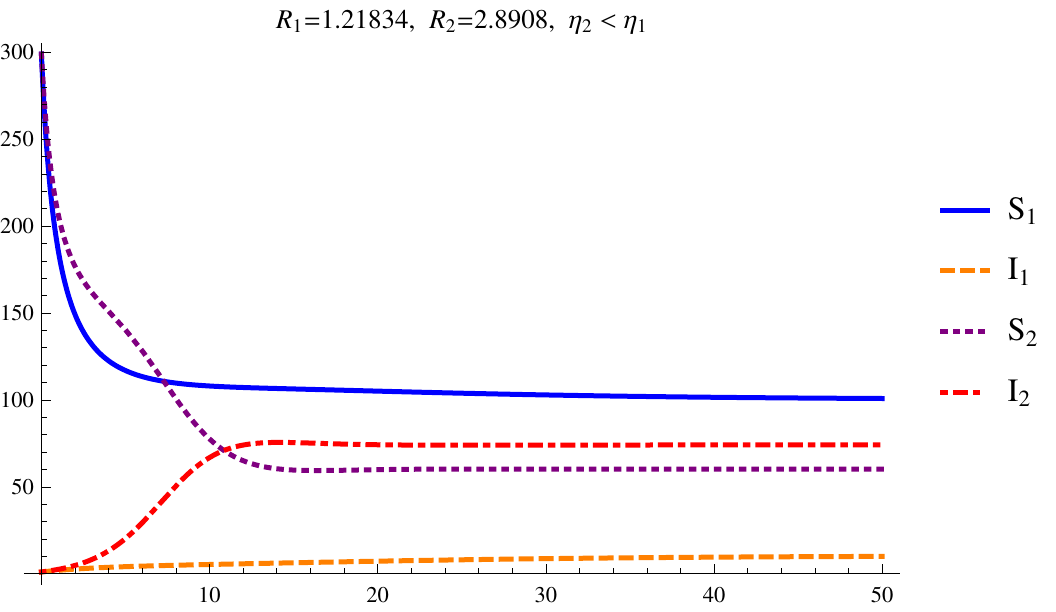}
\end{subfigure}%
\begin{subfigure}{.5\textwidth}
  \centering
  \includegraphics[scale = 0.7]{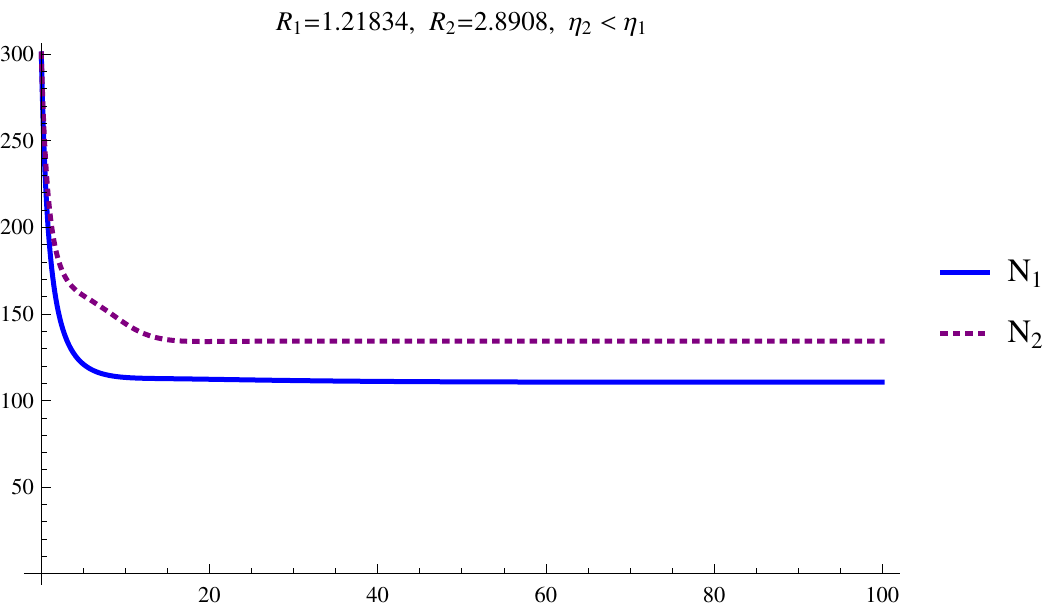}
\end{subfigure}
\caption{$\omega_{21} = 2 \cdot \omega_{12}$ and $\eta_1 = 3 \cdot \eta_2$, other parameters fixed}
\label{fig:scenario4}
\end{figure}
Although species 1 is the dominant predator, because it's human-related mortality rate is three times as high (perhaps because it is hunted more) the subordinate predator's population will grow and stabilize at larger numbers than that of the dominant predator.

\section{Conclusions}
In this study we analyzed a four-dimensional system of differential equations that modeled interference competition with an infectious disease and human-related mortality. We developed the model in order to understand the dynamics of competitive predators that both face disease and human-related mortality. In our model we have shown the stability of the three boundary equilibrium, including the trivial equilibrium and the two one-host disease free equilibrium. The parameter space for the one-host endemic equilibriums exist is defined, however we were unable to define the region algebraically for when they are stable. However, through numerical simulations we were able to show that both of these equilibrium points are locally asymptotically stable in some parameter space. The parameter space for when both species can coexist, with and without the disease, is difficult to define algebraically. But we have shown that there is such parameter space for when the coexistence endemic equilibrium and the coexistence disease free equilibrium exist, and that they are locally asymptotically stable. 

Moreover, we have shown that a subordinate predator can play a role in controlling the population size of the more dominant predator. In Scenario 1, where the carrying capacity for the subordinate species $(k_2)$ is larger than the more dominant species $(k_1)$ shown in Figure \ref{fig:scenario1}, we see that because the environment can support more of the subordinate species that their population size gives them a competitive advantage over the dominant species. In Scenario 2, as shown in Figure \ref{fig:scenario2} when the dominant predator's transmission coefficient $\beta_1$ is larger, if for example they are a more social species or if their immune system is not as strong as the other species', then the subordinate predator's population will grow larger and help regulate the other species both through competition and the size of the population. 

In the other scenarios we ran, shown in Figure \ref{fig:scenario3} and Figure \ref{fig:scenario4}, we needed to increase the human-related mortality and the disease related deaths of the dominant species to three times that of the subordinate species for it to play a role in determining which population size was larger, and which species wins the competition. However, the outcomes are very dependent on the initial conditions and the parameters, and take these simulations as definitive answers to the outcome of competition in every case. But in some cases we can see that disease death and human-related mortality can shift the dynamics of the system so that the subordinate predator wins the competition. Further work will be needed to determine the stability of the entire system, and should include the addition of social structure. In, for example, the case of the wolves and coyotes of Yellowstone National Park, the more social and dominant creature, the wolf, will have a disadvantage due to higher disease transmission because of larger pack sizes. Moreover, if the population sizes are small, we could also consider a spatially explicit model to more accurately reflect the dynamics of the disease. The eco-epidemiological framework here serves as a basis to further explore the dynamics of competitive predators under different environmental influences. 

\section*{Acknowledgments} 
We would like to thank Dr. Carlos Castillo-Chavez, Executive Director of the Mathematical and Theoretical Biology Institute (MTBI), for giving us this opportunity to participate in this research program. We would also like to thank Co-Executive Summer Directors Dr.~Omayra Ortega and Dr.~Baojun Song for their efforts in planning and executing the day to day activities of MTBI. I would like to give a special thanks to Komi Messan and Juan Renova for their help and patience. This research was conducted in MTBI at the Simon A. Levin Mathematical, Computational and Modeling Sciences Center (SAL MCMSC) at Arizona State University (ASU).  This project has been partially supported by grants from the National Science Foundation (DMS-1263374 and DUE-1101782), the National Security Agency (H98230-14-1-0157), the Office of the President of ASU, and the Office of the Provost of ASU.

\newpage
\bibliography{mybib_project}

\newpage
\section{Appendix}
\subsection{Stability of the Two-Host Disease Free Equilibrium} \label{app:one-host_DFE}
The two-host disease free equilibrium is given by

$$E_{5}=\left(\frac{\alpha_1k_2\omega_{12}(\eta_2-\alpha_2)+\alpha_2k_1(\alpha_1-\eta_1)}{\alpha_2\alpha_1(1-\omega_{21}\omega_{12})},0,\frac{\alpha_2k_1\omega_{21}(\eta_1-\alpha_1)+\alpha_1k_2(\alpha_2-\eta_2)}{\alpha_2\alpha_1(1-\omega_{21}\omega_{12})},0\right) $$

\noindent and will exist if

$$\eta_1 <\alpha_1\text{ and }\eta_2 < \alpha_2 \text{ and } \omega_{12}\omega_{21} < 1,$$ \\

\noindent which means that the intrinsic growth rate of each population is greater than the respective rates at which they are hunted, and if the competition of the species is relatively low, both species cannot be using extreme amount of resources. Taking the Jacobian, $J$, of the system and evaluating it at $E_{5}$, we get
{\tiny
$$\left[ \begin {array}{cccc} -{\frac {{\it S^{*}_2}\,\alpha_{{1}}\omega_{{
12}}+2\,{\it S^{*}_1}\,\alpha_{{1}}-\alpha_{{1}}k_{{1}}+\eta_{{1}}k_{{1}}
}{k_{{1}}}}&-{\frac {{\it S^{*}_2}\,\alpha_{{1}}\omega_{{12}}+{\it S^{*}_1}\,
\beta_{{1}}k_{{1}}+2\,{\it S^{*}_1}\,\alpha_{{1}}-\alpha_{{1}}k_{{1}}-
\gamma_{{1}}k_{{1}}}{k_{{1}}}}&-{\frac {{\it S^{*}_1}\,\alpha_{{1}}\omega_
{{12}}}{k_{{1}}}}&-{\frac {{\it S^{*}_1}\,\alpha_{{1}}\omega_{{12}}}{k_{{1
}}}}\\ \noalign{\medskip}0&{\it S^{*}_1}\,\beta_{{1}}-G_{{1}}&0&0
\\ \noalign{\medskip}-{\frac {{\it S^{*}_2}\,\alpha_{{2}}\omega_{{21}}}{k_
{{2}}}}&-{\frac {{\it S^{*}_2}\,\alpha_{{2}}\omega_{{21}}}{k_{{2}}}}&-{
\frac {{\it S^{*}_1}\,\alpha_{{2}}\omega_{{21}}+2\,{\it S^{*}_2}\,\alpha_{{2}}
-\alpha_{{2}}k_{{2}}+\eta_{{2}}k_{{2}}}{k_{{2}}}}&-{\frac {{\it S^{*}_2}\,
\beta_{{2}}k_{{2}}+{\it S^{*}_1}\,\alpha_{{2}}\omega_{{21}}+2\,{\it S^{*}_2}\,
\alpha_{{2}}-\alpha_{{2}}k_{{2}}-\gamma_{{2}}k_{{2}}}{k_{{2}}}}
\\ \noalign{\medskip}0&0&0&{\it S^{*}_2}\,\beta_{{2}}-G_{{2}}\end {array} \right]$$
}

\noindent The determinant of J is

$$det(J)=(J_{22}-\lambda)(J_{44}-\lambda)({\lambda}^{2}+\left(-{\it J_{11}}-{\it J_{33}}\right)\lambda+{\it J_{11}}\,{\it J_{33}}-{\it J_{31}}\,{\it J_{13}})$$

\noindent Since the eigenvalues of $J$ need to be negative, we need 

$$S_{1}^{*}<\frac{G_{1}}{\beta_1}$$

\noindent and

$$S_{2}^{*}<\frac{G_{2}}{\beta_2}$$

\noindent for our first two conditions for stability. Rearranging these inequalities we get that 

$$R_1 < R_2 \left( \frac{G_2}{\beta_2} \frac{\beta_1}{G_1} \right) \omega_{12} \text{ and }R_2 < R_1 \left( \frac{G_1}{\beta_1} \frac{\beta_2}{G_2} \right) \omega_{21}$$

\noindent so that 

$$R_2 \left( \frac{G_2}{\beta_2} \frac{\beta_1}{G_1} \right) \omega_{21} < R_1 < R_2 \left( \frac{G_2}{\beta_2} \frac{\beta_1}{G_1} \right) \omega_{12}$$

This restricts the average number of secondary infections for each species so that it is not bigger than a fraction of the other. Now we use the Routh-Hurwitz criterion on the quadratic 

$$({\lambda}^{2}+\left(-{\it J_{11}}-{\it J_{33}}\right)\lambda+{\it J_{11}}\,{\it J_{33}}-{\it J_{31}}\,{\it J_{13}})$$

\noindent derived from the Jacobian. The Routh-Hurwitz criterion for a quadratic equation states that 

$$J_{11}+J_{33}<0 \text{ and } \left({\it J_{11}}\,{\it J_{33}}-{\it J_{31}}\,{\it J_{13}} \right) >0 $$

\noindent First, we need $J_{11}+J_{33}<0$, i.e. 
 
$${\alpha_1\left(1-\frac{2S_{1}^{*}+\omega_{12}S_{2}^{*}}{k_1}\right)-\eta_1}+{\alpha_2\left(1-\frac{2S_{2}^{*}+\omega_{21}S_{1}^{*}}{k_2}\right)-\eta_2}<0$$

\noindent Rearranging this inequality, we get 

$${(\alpha_1-\eta_1)-\alpha_1 \left(\frac{2S_{1}^{*}+\omega_{12}S_{2}^{*}}{k_1}
\right)}+{(\alpha_2-\eta_2)-\alpha_2\left(\frac{2S_{2}^{*}+\omega_{21}S_{1}^{*}}{k_2}\right)}<0$$

\noindent We also need for $J_{11}J_{33} - J_{31}J_{13} >0$, i.e.

$$\left({\alpha_1\left(1-\frac{2S_{1}^{*}+\omega_{12}S_{2}^{*}}{k_1}\right)-\eta_1}\right)\left({\alpha_2\left(1-\frac{2S_{2}^{*}+\omega_{21}S_{1}^{*}}{k_2}\right)-\eta_2}\right)-\left({\frac{{\it S_{1}^{*}}\,\alpha_{{w}}\omega_{{\it 12}}}{k_{{w}}}}\right)\left({\frac{{\it S_{2}^{*}}\,\alpha_{{c}}\omega_{{\it 21}}}{k_{{c}}}}\right)>0$$

$$\left({(\alpha_1-\eta_1)-\alpha_1\left(\frac{2S_{1}^{*}+\omega_{12}S_{2}^{*}}{k_1}\right)}\right)\left({(\alpha_2-\eta_2)-\alpha_2\left(\frac{2S_{2}^{*}+\omega_{21}S_{1}^{*}}{k_2}\right)}\right)-\left({\frac{{\it S_{1}^{*}}\,\alpha_{{w}}\omega_{{\it 12}}}{k_{{w}}}}\right)\left({\frac{{\it S_{2}^{*}}\,\alpha_{{c}}\omega_{{\it 21}}}{k_{{c}}}}\right)>0$$

\noindent Expanding the inequality above will give us 

{\tiny
$$(\alpha_1-\eta_1)(\alpha_2-\eta_2) - \alpha_2(\alpha_1 - \eta_1)\left(\frac{2S^{*}_2 + \omega_{21} S^{*}_1}{k_2}\right) - \alpha_1(\alpha_2-\eta_2)\left(\frac{2S^{*}_1 + \omega_{12} S^{*}_2}{k_1}\right) + \alpha_2 \alpha_1 \left(\frac{2S^{*}_2 + \omega_{21} S^{*}_1}{k_2}\right) \left(\frac{2S^{*}_1 + \omega_{12} S^{*}_2}{k_1}\right) > \left(\frac{S^{*}_1 \alpha_1 \omega_{12}}{k_1} \frac{S^{*}_2 \alpha_2 \omega_{21}}{k_2} \right)$$
}

{
$$(\alpha_1-\eta_1)(\alpha_2-\eta_2) - \alpha_2(\alpha_1 - \eta_1)\left(\frac{2S^{*}_2 + \omega_{21} S^{*}_1}{k_2}\right) - \alpha_1(\alpha_2-\eta_2)\left(\frac{2S^{*}_1 + \omega_{12} S^{*}_2}{k_1}\right) + \alpha_2 \alpha_1 \left(\frac{2S^{*}_2}{k_2}\right) \left(\frac{2S^{*}_1}{k_1}\right) > 0$$
}

\noindent Because the conditions for when $J_{11}+J_{33}<0$ and $J_{11}J_{33} - J_{31}J_{13} >0$ are both algebraically complex, and the overlap of the conditions is indeterminable due to the number of parameters, we will run numerical simulations to determine the stability of the two-host DFE.

\subsection{Unique Positive Solution for One-Host Endemic Equilibrium} \label{unique_pos}
There are two one-host endemic equilibrium points, one for each of the species in our system. These two equilibrium points are as follows:
\begin{align*}
E_3 &= \left(\frac{G_1}{\beta_1},I^{*}_1,0,0 \right)  \\
E_4 &= \left(0,0,\frac{G_2}{\beta_2},I^{*}_2 \right)  
\end{align*}

\noindent where

\begin{align}
A &= 1/2\,{\frac {-k_{{1}} \left( G_{1}-\alpha_{{1}}-\gamma_{{1}} \right) 
\beta_{{1}}-2\,G_{1}\,\alpha_{{1}} \pm \sqrt {-4\,\beta_{{1}} \left( -1/4
\,k_{{1}} \left( G_{1}-\alpha_{{1}}-\gamma_{{1}} \right) ^{2}\beta_{{1
}}+G_{1}\,\alpha_{{1}} \left( -\delta_1 \right)  
\right) k_{{1}}}}{\alpha_{{1}}\beta_{{1}}}} \label{A}\\
B &= 1/2\,{\frac {-k_{{2}} \left( G_{2}-\alpha_{{2}}-\gamma_{{2}} \right) 
\beta_{{2}}-2\,G_{2}\,\alpha_{{2}} \pm \sqrt {-4\,\beta_{{2}} \left( -1/4
\,k_{{2}} \left( G_{2}-\alpha_{{2}}-\gamma_{{2}} \right) ^{2}\beta_{{2
}}+G_{2}\,\alpha_{{2}} \left( -\delta_2 \right) 
\right) k_{{2}}}}{\alpha_{{2}}\beta_{{2}}}} \label{B}
\end{align}

\noindent and $I^{*}_1$ is the positive value of $A$ and $I^{*}_2$ is the positive value of $B$.

We will prove that only one of the values of $A$ and one of the values of $B$ are positive, and consequently there are only two one-host endemic equilibrium points. Because of the symmetry of the expressions for A and B, we will prove a single positive value for A. 

\noindent We note that A is the solution for the quadratic 

\begin{align}
-{\frac {\alpha_{{1}}{I_{{1}}}^{2}}{k_{{1}}}}+ \left( -{\frac {G_{1}\,
\alpha_{{1}}}{k_{{1}}\beta_{{1}}}}+\alpha_{{1}} \left( 1-{\frac {G_{1}
}{k_{{1}}\beta_{{1}}}} \right) +\gamma_{{1}}-G_{1} \right) I_{{1}}+{
\frac {G_{1}\,\alpha_{{w}}}{\beta_{{1}}} \left( 1-{\frac {G_{1}}{k_{{1
}}\beta_{{1}}}} \right) }-{\frac {\eta_1\,G_{1}}{\beta_{{1}}}} \label{quad_A}
\end{align}

\noindent  and so we let 

\begin{align*}
a &= -{\frac {\alpha_{{1}}{I_{{1}}}^{2}}{k_{{1}}}} \\
b &= \left( -{\frac {G_{1}\,
\alpha_{{1}}}{k_{{1}}\beta_{{1}}}}+\alpha_{{1}} \left( 1-{\frac{G_{1}
}{k_{{1}}\beta_{{1}}}} \right) +\gamma_{{1}}-G_{1} \right) \\
c &= {\frac {G_{1}\,\alpha_{{1}}}{\beta_{{1}}} \left( 1-{\frac {G_{1}}{k_{{1
}}\beta_{{1}}}} \right) }-{\frac {\eta_1\,G_{1}}{\beta_{{1}}}}
\end{align*}

We can see from equation \eqref{A} that the determinant of the quadratic in equation \eqref{quad_A} will always be positive, and since all parameters are positive we know that $a<0$ for all parameters. Therefore by determining the signs of $b$ and $c$ we can come to a conclusion about the number of positive roots that exist. Suppose $b > 0$. Then 

\begin{align*}
-{\frac {G_{1}\, \alpha_{{1}}}{k_{{1}}\beta_{{1}}}}+\alpha_{{1}} \left( 1-{\frac{G_{1}}{k_{{1}}\beta_{{1}}}} \right) +\gamma_{{1}}-G_{1} &> 0 \\
-{2\frac {G_{1}\, \alpha_{{1}}}{k_{{1}}\beta_{{1}}}}+\alpha_{{1}} +\gamma_{{1}}-G_{1} &> 0 \\
\end{align*}

\noindent Since $G_1 = \eta_1 + \gamma_1 + \delta_1$, we use substitution to obtain

\begin{align*}
-2 \alpha_1 G_1 + \beta_1 k_1 \alpha_1 - \beta_1 k_1 (\delta_1 + \eta_1) &> 0 \\
-2 \alpha_1 G_1 - \beta_1 k_1 \alpha_1 \left(1 - \frac{\delta_1 + \eta_1}{\alpha_1}\right) &> 0 \\
-2 \alpha_1 G_1 - \beta_1 k_1 \alpha_1 \left(1 - \frac{\eta_1}{\alpha_1}\right) - \beta_1 k_1 \delta_1 &> 0 \\
-2 G_1 - \beta_1 k_1 \left(1 - \frac{\eta_1}{\alpha_1}\right) - \frac{\beta_1 k_1 \delta_1}{\alpha_1} &> 0 \\
-2 G_1 - \frac{\beta_1 k_1}{\alpha_1} \left(\alpha_1 - \eta_1 \right) - \frac{\beta_1 k_1 \delta_1}{\alpha_1} &> 0 \\
-2 - \frac{\beta_1 k_1}{\alpha_1 G_1} \left(\alpha_1 - \eta_1 \right) - \frac{\beta_1 k_1 \delta_1}{\alpha_1 G_1} &> 0 \\
\end{align*}

\noindent Since $R_1 = \frac{\beta_1 k_1}{\alpha_1 G_1} \left(\alpha_1 - \eta_1 \right)$ we get

\begin{align*}
-2 + R_1 - \frac{\beta_1 k_1 \delta_1}{\alpha_1 G_1} &> 0 \\
R_1 &> 2 + \frac{\beta_1 k_1 \delta_1}{\alpha_1 G_1} 
\end{align*}

\noindent Now suppose $c < 0$, i.e.

\begin{align*}
{\frac {G_{1}\,\alpha_{{1}}}{\beta_{{1}}} \left( 1-{\frac {G_{1}}{k_{{1
}}\beta_{{1}}}} \right) }-{\frac {\eta_1\,G_{1}}{\beta_{{1}}}} &< 0 \\
(\alpha_1 - \eta_1) - \frac{\alpha_1 G_1}{\beta_1 k_1} &< 0 \\
k_1(\alpha_1 - \eta_1) - \frac{\alpha_1 G_1}{\beta_1} &< 0 \\
\frac{\beta_1 k_1(\alpha_1 - \eta_1)}{G_1 \alpha_1} - 1 &< 0 \\
R_1 &< 1
\end{align*}

\noindent Using Vieta's relations for quadratic equations, we know that the sum of the roots, say $x$ and $y$, of a quadratic equation is $x + y = -b/a$ and the product of the roots is $xy = c/a$. In order to have two positive roots then we must have that the sum of the roots be positive and the product of the the roots to be positive, i.e. $-b/a > 0$ and $c/a > 0$. Since $a < 0$ is always true, we need for $c < 0$ to satisfy the first condition. Therefore we need $R_1 < 1$. In order for $-b/a > 0$ we need for $b<0$. However, this is a contradiction since then $R_1 > 2 + \frac{\beta_1 k_1 \delta_1}{\alpha_1 G_1}$, hence we cannot have two positive roots.   

Now suppose both roots were negative. Then according to Vieta's theorem we would need the sum of the roots to be negative and the product of the roots to be positive, i.e. $-b/a < 0$  and $c/a > 0$. Since $a < 0$, then we need $b<0$. We can see that the same contradiction is reached. 

Since the determinant is positive, there are no complex roots. Since both roots are not positive, and both roots are not negative we must have one positive root and one negative root. Therefore, there is at most one positive solution, and at most one biologically relevant equilibrium point.\\

\noindent The proof for B follows. \qed

\subsection{Numerical Simulations to Show Stability}
      \begin{figure}[H]
       \centering
       \includegraphics[scale = 1.35]{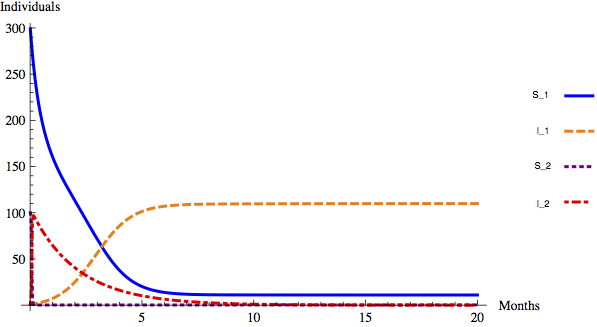} 
       \caption{One-Host Endemic Equilibrium:  Only Species 1 Survives with Infection}
       \end{figure}

     \begin{figure}[H]
       \centering
       \includegraphics[scale = 1.35]{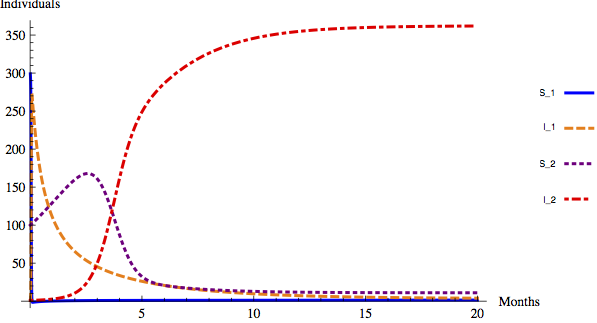} 
       \caption{One-Host Endemic Equilibrium: Only Species 2 Survives with Infection}
       \end{figure}

     \begin{figure}[H]
       \centering
       \includegraphics[scale = 1.35]{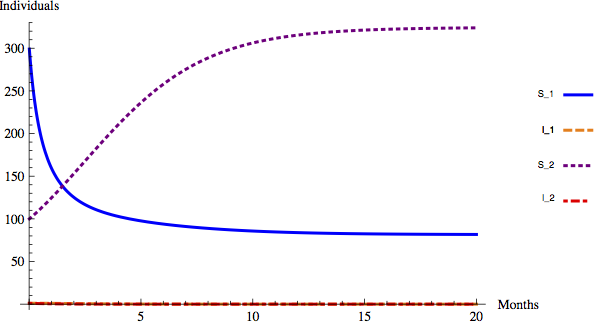} 
       \caption{Coexistence Disease Free Equilibrium: Both Species Survive without Infection}
       \end{figure}

     \begin{figure}[H]
       \centering
       \includegraphics[scale = 1.35]{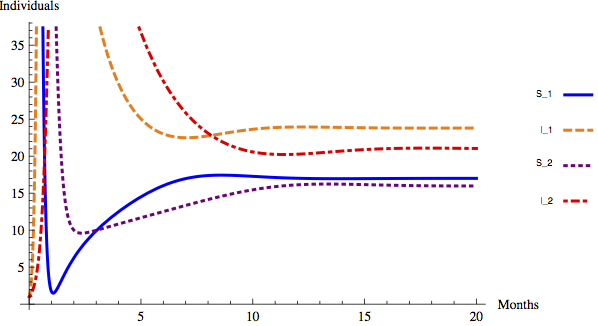} 
       \caption{Coexistence Endemic Equilibrium: Both Species Survive with Infection}
       \end{figure}
\end{document}